\documentclass[aps,pra,superscriptaddress,amsmath,amssymb,preprintnumbers,floatfix,showpacs,11pt]{revtex4}

\usepackage{amssymb}
\usepackage{epsfig}
\usepackage{subfigure}
\usepackage{graphicx}
\begin{document}

\title{Squeezing properties of the Kerr-down conversion system
}
\author{ Faisal A. A. El-Orany}
 \affiliation{
Department of Mathematics  and Computer Science, Faculty of
Science, Suez Canal University, 41522 Ismailia, Egypt}

\author{ M. Sebawe Abdalla}
 \affiliation{Mathematics Department, College of Science, King Saud
University, P.O. Box 2455, Riyadh 11451, Saudi Arabia}

\author{ and J. Pe\v {r}ina}
 \affiliation{Department of Optics and Joint Laboratory of
Optics, Palack\'{y} University, 17. listopadu 50,
 772 07 Olomouc, Czech Republic}


\begin{abstract}
In this Letter  we describe a new two-mode  system, which consists
of Kerr-like medium and down conversion process, called the
Kerr-down conversion
system. Under a certain condition we can obtain an exact solution
of the
dynamical equations of motion. For this system we investigate
different kinds of quadrature squeezing, e.g., single-mode,
two-mode and sum-squeezing. Also we give a more general definition
of the principal squeezing. We show that the amounts of
nonclassical effects produced by the Kerr-like and
down-conversion processes separately are greater than those obtained
from
the Kerr-down conversion system where both the processes are in
competition.

\end{abstract}

 \pacs{42.50.Dv,42.50.-p} \maketitle

\section{Introduction}
There are three important processes in nonlinear optics, namely,
up-conversion, down-conversion and Kerr-like process. These
processes took most of attention in the field of quantum optics.
The up-conversion process cannot generate nonclassical effects,
however, it  switches energy between modes, e.g., signal and idler
\cite{mol}. In this regard the up-conversion process is mainly
used in the directional couplers to transfer data between
waveguides \cite{per1}. The down-conversion, which is usually
named in the literature as parametric amplifier, can generate
nonclassical effects. Precisely, the degenerate  and nondegenerate
parametric amplifiers perfectly generate single-mode \cite{mol1}
and two-mode \cite{sqz1} squeezed states, respectively. Actually,
squeezed states have been applied in the quantum information
\cite{hil}. The single-mode state obtained from the nondegenerate
parametric amplifier is  'super-classical' in the sense that the
evolution of the system broadens the single-mode $P$ distribution
as a result of the spontaneous pump photon decays \cite{faisal1}.
This behavior was named later self-decoherence \cite{faisal2}.
Moreover, the parametric amplifier has been used in the
observation of the interference effects. For instance, the
fourth-order interference effects arise when pairs of photons
produced in parametric amplifier are injected into Michelson
interferometers \cite{mich}. Additionally, the second-order
interference is observed in the superposition of signal photons
from two coherently pumped parametric amplifiers when the paths of
the idler photons are aligned \cite{path}. Now we  draw the
attention to the interaction of light with the Kerr-like medium.
This type of interaction is representative by generating  cat
states, namely, the Yurke-Stoler state \cite{ker2}. The
Yurke-Stoler state can generate nonclassical squeezing in spite of
the fact that the
 photon-number distribution is Poissonian.
The Kerr-like medium has been intensively studied for various
 kinds of the initial states aiming to obtain nonclassical
  squeezed light , e.g.,
\cite{ker1}. Such studies are encouraged by the possible
observation of the large values of the third-order optical
nonlinearities in, e.g., the organic polymers \cite{exp2}.

The competition between the down-conversion (the Kerr-like medium)
and up-conversion processes is of interest from the theoretical
and experimental points of views, e.g., in the three-mode
interaction \cite{inter1} and in the nonlinear directional
couplers \cite{dir,coup}. Nevertheless, the competition between
the Kerr-medium and
 nondegenerate parametric amplifier--as far as we know--has not been
treated yet.
 Thus in this Letter we investigate this system, i.e., the
Kerr-down conversion system. For this system the exact solution
can only be obtained under certain condition, as we shall see. As
we mentioned above the Kerr medium and the down-conversion are
important processes, so that the connection between them is
important, too. Moreover, this system enables us to investigate
the influence of the Kerr medium on the down-conversion and vice
versa. For this  system we investigate  different types of
quadrature squeezing such as single-mode, two-mode and sum
squeezing. Also we develop the notion of the principal squeezing
\cite{prin} for any type of the quadrature squeezing. It is worth
reminding that the squeezed light
 can be measured in the
homodyne detector where the signal is superimposed on the
strong coherent beam of the local oscillator.
 Generally, we show that the nonclassical
effects generated by this system are smaller   than those obtained
from the individual processes, i.e., each process  destroys  the
nonclassical effects generated  by the other one. This contradicts
with what one could possibly expect, i.e., that the combination of
these
processes will increase the nonclassical effects.
We treat this problem in the following order:
In section 2 we give the Hamiltonian
model of the system and the basic equations and relations. In
section 3 we discuss the corresponding results.

\section{Basic relations and equations}

In this section we give the Hamiltonian model and solve the
associated equations of motion. Also we write down  the
definitions of the quadrature squeezing and develop general
definition to the principal squeezing.
 The Hamiltonian model for the Kerr-down conversion system  can be
expressed as
\begin{equation}
\frac{\hat{H}(t)}{\hbar}=\sum\limits_{j=1}^{2}[\omega_j\hat{a}_j^{\dagger}\hat{a}_j
+\chi_j\hat{a}_j^{\dagger
2}\hat{a}_j^{2}]+\bar{\chi}\hat{a}_1^{\dagger}\hat{a}_1
\hat{a}_2^{\dagger}\hat{a}_2-ik [\hat{a}_1\hat{a}_2\exp(i\omega t)
-\hat{a}_1^{\dagger}\hat{a}_2^{\dagger}\exp(-i\omega
t)],\label{h1}
\end{equation}
where the waves are described by annihilation operators
$\hat{a}_j$ and by frequencies $\omega_j\quad
(j=1,2)$ of the first and second
mode, respectively.  The coupling constants $\chi_j$ and
$\overline{\chi }$ are proportional to the third-order
susceptibility $\chi^{(3)}$ and are responsible correspondingly
for the self-action and cross-action processes of the modes.
The coupling constant $k$ is real and $\omega=\omega_1+\omega_2$.
The Hamiltonian (\ref{h1}) can be obtained in terms of two modes
which are
interacting via a multi-layer nonlinear crystal comprising from
the Kerr medium and down conversion medium. This could be possible
with respect to the
progress of preparation of new nonlinear crystals and
improved laser sources
\cite{exp2,{cry}}.

The exact solution for the equations of motion for (\ref{h1})
can be obtained under the assumption that
$\overline{\chi}=-2\chi_1=-2\chi_2$. In
this case $\hat{N}=\hat{n}_1-\hat{n}_2$, where
$\hat{n}_j=\hat{a}_j^{\dagger}\hat{a}_j$, is a constant of motion
and  the solution takes the form:

\begin{eqnarray}
\begin{array}{lr}
\hat{A}_{1}(t)=\exp (-2i\overline{\chi} \hat{N}t)\Bigl\{ \hat{a}%
_{1}(0)C+\hat{a}_{2}^{\dagger}(0)S\Bigr\} , \\
\\
\hat{A}_{2}(t)=\exp (2i\overline{\chi} \hat{N}t)
\Bigl\{ \hat{a}%
_{2}(0)C+\hat{a}_{1}^{\dagger}(0)S\Bigr\}, \label{10}
\end{array}
\end{eqnarray}
where $\hat{A}_{j}(t)=\exp (i\omega_j t)
 \hat{a}_j$ and
  we have used the abbreviations
\begin{equation}
C=\cosh (k t),\quad S=\sinh (k t). \label{h2}
\end{equation}
It is evident that the expressions (\ref{10}) include the
amplification and periodical features of the down-conversion and
Kerr-like processes, respectively. The Kerr process is represented
by the non-trivial phase factor, which  plays an essential role in
occurrence of the  nonclassical effects.

As we investigate various types of squeezing, i.e. single-mode,
two-mode, and sum squeezing,
 we define two general quadratures as
$\hat{X}=\frac{1}{2}(\hat{B}+\hat{B}^{\dagger})$ and
 $\hat{Y}=\frac{1}{2i}(\hat{B}-\hat{B}^{\dagger})$,  which satisfy
 the commutation rule $[\hat{X},\hat{Y}]=\frac{\hat{D}}{2i}$. The
 operators $\hat{B}$ and $\hat{D}$ will be specified in the
text. Squeezing factors associated with the $\hat{X}$ and
$\hat{Y}$ can be expressed, respectively, as
\begin{eqnarray}
\begin{array}{rl}
 F =\frac{1}{|\langle\hat{D}\rangle|}\left[ 2{\rm
Re}\langle\hat{B}^2\rangle+2\langle\hat{B}^{\dagger}\hat{B}\rangle+
\langle\hat{D}\rangle-|\langle\hat{D}\rangle|-4[{\rm
Re}\langle\hat{B}\rangle]^2 \right], \\
\\
G
=\frac{1}{|\langle\hat{D}\rangle|}\left[2\langle\hat{B}^{\dagger}\hat{B}\rangle-
2{\rm Re}\langle\hat{B}^2\rangle+
\langle\hat{D}\rangle-|\langle\hat{D}\rangle|-4[{\rm
Im}\langle\hat{B}\rangle]^2 \right]. \label{ol5}
\end{array}
\end{eqnarray}

We conclude this section by shedding  light on  the principal
squeezing \cite{prin}. For principal squeezing
 the quadrature is defined as:

\begin{equation}
\hat{X}_\phi=\frac{1}{2}[ \hat{B}\exp(-i\phi)+
\hat{B}^{\dagger}\exp(i\phi)].  \label{ol1}
\end{equation}
The value of the angle $\phi$ can be controlled by the homodyne
detector to obtain the maximum amount of squeezing. The squeezing
factor related to (\ref{ol1}) is
\begin{equation}
 V_\phi =\frac{1}{|\langle\hat{D}\rangle|}\left[
4\langle(\triangle \hat{X}_\phi)^2\rangle -|\langle\hat{D}\rangle|
\right]. \label{ol2}
\end{equation}
By evaluating the extreme values for (\ref{ol2}) with respect to
$\phi$, one can obtain general form for the principal squeezing as

\begin{equation}
 V =\frac{1}{|\langle\hat{D}\rangle|}\left[\langle\hat{D}\rangle+
2\langle\hat{B}^{\dagger}\hat{B}\rangle-(2\langle\hat{B}\rangle
\langle\hat{B}^{\dagger}\rangle+ |\langle\hat{D}\rangle|)-2 \Bigl|
\langle\hat{B}^2\rangle-\langle\hat{B}\rangle^2\Bigr|\right].
\label{ol3}
\end{equation}
The definition (\ref{ol3})  is more general than that given in
 \cite{prin} in which the principal squeezing
 definition  is derived only for the
 single-mode and two-mode cases through lengthy calculations.

\section{Discussion of the results}
We use the results given in the previous section to investigate
the quadrature squeezing for the system under consideration. We
focus the attention on the single-mode, two-mode and sum
 squeezing. We assume that the two modes are initially prepared in the coherent states
 $|\alpha_1,\alpha_2\rangle$ with real amplitudes $\alpha_j$.
 Also we shed light on the evolution of
the principal squeezing for each type. This will be done in the
following.
\subsection{Single-mode squeezing}
As is well known that
 the Kerr-like medium can provide nonclassical squeezing \cite{ker1},
however, the nondegenerate parametric amplifier cannot generate
single-mode squeezing as a result of the self-decoherence
\cite{faisal2}. In the present system we found that the Kerr-like
medium is responsible for generating periodical single-mode
squeezing provided that $k$ is very small. This means that the
competition between the two processes, i.e. Kerr-like medium and
down-conversion, may destroy the nonclassical effects inherited in
the individual ones. We show this fact for the first mode.
 In this case we have $\hat{B}(t)=\hat{A}_1(t), \hat{D}=1$ and
 the squeezing factors have the forms:
\begin{figure}
  \vspace{0cm}
\centerline{\epsfxsize=7cm \epsfbox{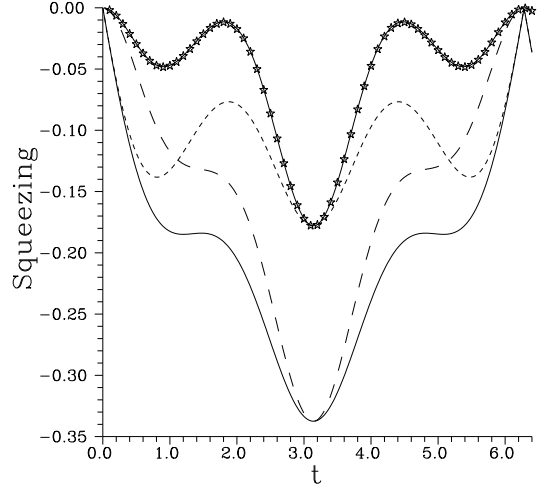}}
 \vspace{0.2cm} \caption{
The principal squeezing and the squeezing factor $F_1^{(1)}(t)$
for the first mode against the interaction time $t$ when $(\chi,
k)=(0.5,0)$ and $(\alpha_1,\alpha_2)=(0.4,0)$ (solid and
long-dashed curves) and $(0.4,0.4)$ (short-dashed and
star-centered curves).}
\end{figure}

\begin{eqnarray}
\begin{array}{lr}
F_1^{(1)}(t)= 2[\alpha_1^2 C^2+2\alpha_1\alpha_2 S
C+S^2(\alpha_2^2+1)]
\\
\\
+2\left[\alpha_1^2 C^2\cos\Theta_++\alpha_2^2
S^2\sin\theta+2\alpha_1\alpha_2 C
S\cos\Theta_-\right]\exp[\epsilon_1\sin^2(2\chi t)]
\\
\\
-4\left[ \alpha_1 C\cos(\epsilon_2\sin(2\chi t)) +\alpha_2
S\cos(2\chi t-\epsilon_2\sin(2\chi t))\right]^2
\exp[\epsilon_1\sin^2(\chi t)],
\\
\\
G_1^{(1)}(t)= 2[\alpha_1^2 C^2+2\alpha_1\alpha_2 S
C+S^2(\alpha_2^2+1)]
\\
\\
-2\left[\alpha_1^2 C^2\cos\Theta_++\alpha_2^2
S^2\sin\theta+2\alpha_1\alpha_2 C
S\cos\Theta_-\right]\exp[\epsilon_1\sin^2(2\chi t)]
\\
\\
-4\left[ \alpha_1 C\sin(\epsilon_2\sin(2\chi t)) -\alpha_2
S\sin(2\chi t-\epsilon_2\sin(2\chi t))\right]^2
\exp[\epsilon_1\sin^2(\chi t)],
 \label{ool3}
 \end{array}
\end{eqnarray}
where the subscript (superscript) $1$ ($1$) means that these
quantities are related to the single-mode case (first mode). Also
in (\ref{ool3}) we have used the following abbreviations:

\begin{eqnarray}
\begin{array}{lr}
\epsilon_1=-2(\alpha_1^2+\alpha_2^2),\qquad
\epsilon_2=\alpha_1^2-\alpha_2^2,
 \\
\\
 \Theta_{\pm}=2\chi t\pm \epsilon\sin(4\chi t),\quad
 \theta=6\chi t - \epsilon\sin(4\chi t).
 \label{olol3}
\end{array}
\end{eqnarray}
One can easily check that  when $(k,\alpha_2)=(0,0)$ the relations
(\ref{ool3}) reduce to those of the single-mode Kerr-like medium.
Now we restrict the analytical discussion to the case
$\alpha_1=\alpha_2$. From (\ref{ool3}) and (\ref{olol3}) one can
easily realized that maximum squeezing occurs when $\chi
t=m\pi/2$, where $m$ is a positive integer. This means that
values of the $\chi$ are responsible for   the periodicity, i.e.
the degree of harmonics, of the squeezing.
 For these values of
the interaction times the expressions (\ref{ool3}) reduce to

\begin{eqnarray}
\begin{array}{lr}
F_1^{(1)}(t)= 2 S^2 -4\alpha_1^2 \exp(\epsilon_1-2k t),
\\
\\
G_1^{(1)}(t)= 4\alpha_1^2 (C+ S)^2+2S^2.
 \label{ool4}
 \end{array}
\end{eqnarray}
 From
(\ref{ool4}) it is evident that squeezing  occurs only in the
$x$-quadrature when $k$ is very small or zero. This indicates
 that the amplification of the down-conversion (related to
increase of quantum noise)
decreases the non-classicality produced by the Kerr-like medium.
 In Fig. 1 we have plotted the principal squeezing and
the quadrature squeezing
 for the first mode when $(\chi,k)=(0.5,0)$.
It is obvious that  the principal squeezing provides amount of
nonclassical squeezing greater  than that of the quadrature
squeezing (compare the solid and long-dashed curves as well as the
short-dashed and the star-centered curves).
 The comparison between the long-dashed
and star-centered curves shows that  the entanglement between
modes  decreases the nonclassical squeezing. Finally, we have
noted that the higher the values of $\alpha_2$ the smaller the
values of the squeezing.

\subsection{Two-mode squeezing}
We study here the two-mode squeezing in which the correlation
between modes starts to play a role. As we mentioned in the
Introduction  that the nondegenerate parametric amplifier produces
perfect nonclassical squeezing in one of the two-mode quadratures.
 Furthermore, the Kerr-like medium can produce nonclassical
two-mode squeezing, too. Thus the overall behavior of the system
could be expected to increase the amount of squeezing.
Nevertheless,  we have noted that this is not so. We show this in
the following. For the two-mode squeezing
$\hat{B}(t)=\hat{A}_1(t)+\hat{A}_2(t), \hat{D}=2$ and the
squeezing factors can be evaluated as:
\begin{figure}
  \centering
  {\includegraphics[width=8cm]{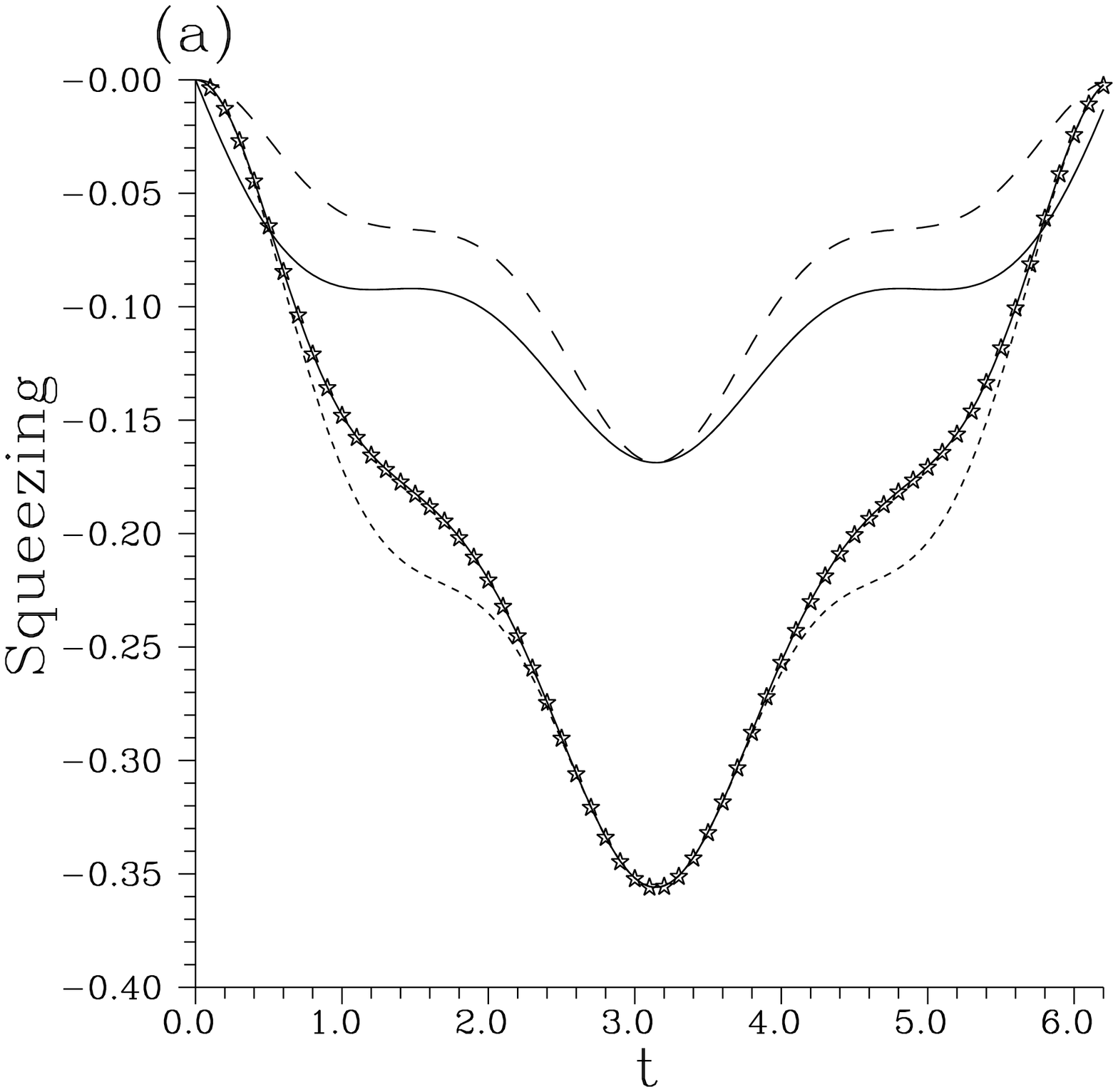}}
 {\includegraphics[width=8cm]{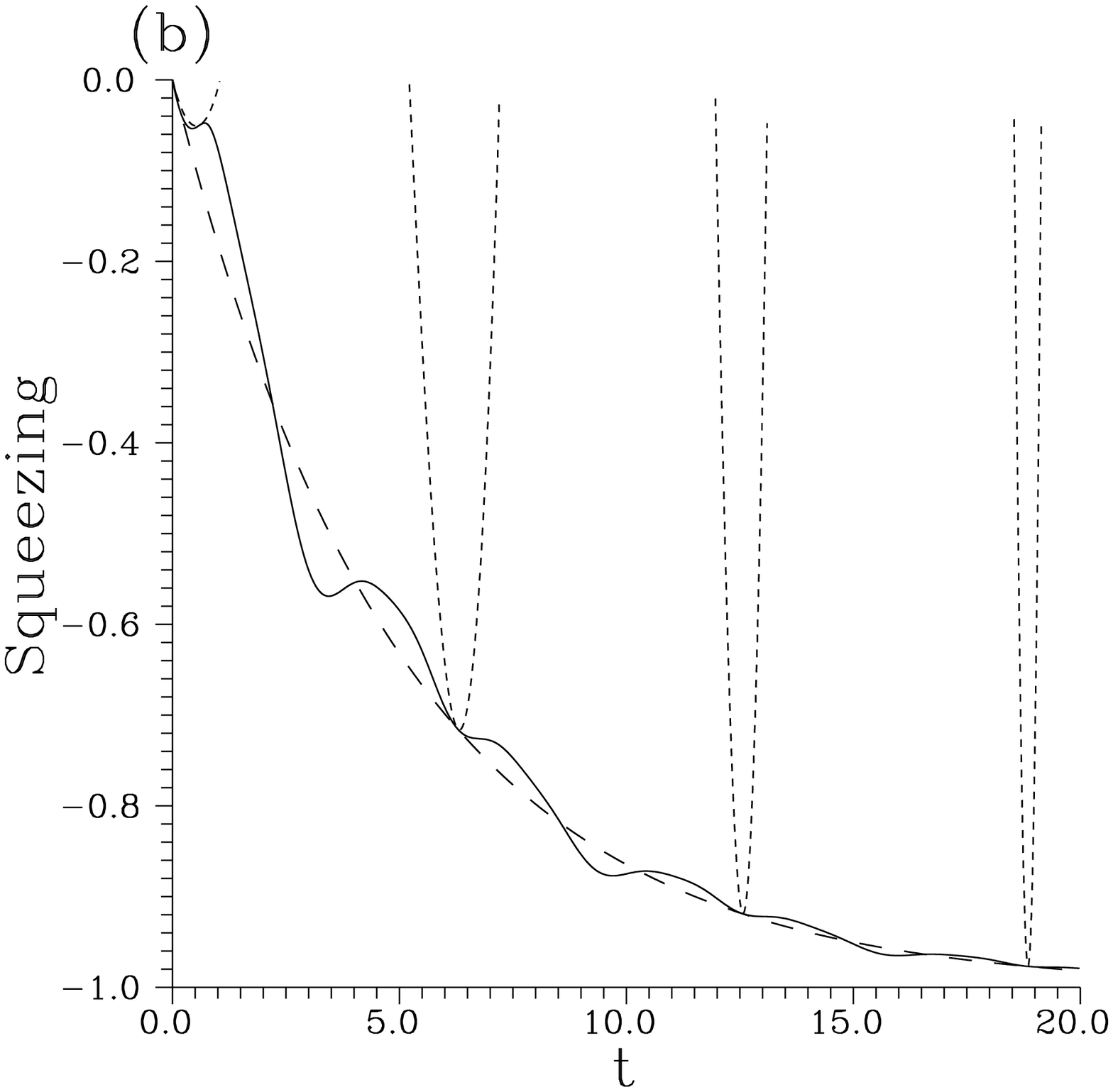}}
  \caption{
The principal squeezing and squeezing factor for the two-mode case
against the interaction time $t$.  (a) For the same situation as
in Fig. 1.  (b) Principal squeezing and  $G_2(t)$ when
$(\alpha_1,\alpha_2)=(0.4,0)$ and $(\chi, k)=(0.5,0.1)$  (solid
and short-dashed curves). The long-dashed curve in (b) is given
for  $(\chi, k)=(0,0.1)$.}
\end{figure}

\begin{eqnarray}
\begin{array}{lr}
F_2(t)=
\frac{1}{2}[ F^{(1)}_1(t)+
F^{(2)}_1(t)]+4[\alpha_1\alpha_2 (S^2+C^2)+S
C(\alpha_1^2+\alpha_2^2+1)]\cos(2\chi t)\\
\\
+4\Bigl\{\alpha_1\alpha_2(C^2+ S^2)\cos(\epsilon_2\sin(4\chi t))
+C
S \alpha_1^2 \cos (4\chi t+\epsilon_2\sin(4\chi t))\\
\\
+ C S \alpha_2^2 \cos (4\chi t-\epsilon_2\sin(4\chi
t))\Bigr\}\exp[\epsilon_1\sin^2(2\chi t)]\\
\\
-8
\left[ \alpha_1 C\cos(\epsilon_2\sin(2\chi t)) +\alpha_2
S\cos(2\chi t-\epsilon_2\sin(2\chi t))\right]
\\
\\
\times \left[ \alpha_2 C\cos(\epsilon_2\sin(2\chi t)) +\alpha_1
S\cos(2\chi t-\epsilon_2\sin(2\chi t))\right]
\exp[2\epsilon_1\sin^2(\chi t)],
\\
\\
G_2(t)= \frac{1}{2}[ G^{(1)}_1(t)+
G^{(2)}_1(t)]-4[\alpha_1\alpha_2 (S^2+C^2)+S
C(\alpha_1^2+\alpha_2^2+1)]\cos(2\chi t)\\
\\
+4\Bigl\{\alpha_1\alpha_2(C^2+ S^2)\cos(\epsilon_2\sin(4\chi t))
+C
S \alpha_1^2 \cos (4\chi t+\epsilon_2\sin(4\chi t))\\
\\
+ C S \alpha_2^2 \cos (4\chi t-\epsilon_2\sin(4\chi
t))\Bigr\}\exp[\epsilon_1\sin^2(2\chi t)]\\
\\
-8 \left[\alpha_1 C\sin(\epsilon_2\sin(2\chi t)) -\alpha_2
S\sin(2\chi t-\epsilon_2\sin(2\chi t))\right]
\\
\\
\times \left[ \alpha_2 C\sin(\epsilon_2\sin(2\chi t)) -\alpha_1
S\sin(2\chi t-\epsilon_2\sin(2\chi t))\right]
\exp[2\epsilon_1\sin^2(\chi t)] ,  \label{fs5}
\end{array}
\end{eqnarray}
where $G_1^{(j)}(t)$ and $F_1^{(j)}(t)$ are the $j$th-single-mode
squeezing factors. We start the discussion by investigating  the
influence of the entanglement on the two-mode Kerr-like medium.
Thus we set $k=0$ in (\ref{fs5}). In this case the maximum amount
of squeezing can periodically occur when $\chi t=m\pi$. At these
values of the interaction time the expressions (\ref{fs5}) reduce
to
\begin{figure}
  \vspace{0cm}
\centerline{\epsfxsize=7cm \epsfbox{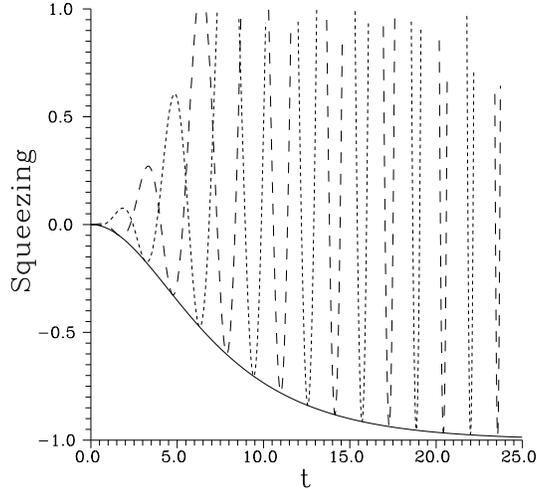} }
 \vspace{0.2cm} \caption{
Sum squeezing  against the interaction time $t$ for
$(\alpha_1,\alpha_2,k)=(0.4,0,0.1)$ and when  $\chi=0$  (solid
 curve-$y$-quadrature), $0.5$ (short-dashed and long-dashed curves for
 $y$- and $x$-quadratures, respectively). Also the solid curve
 represents
 the principal squeezing for the case $\chi=0.5$.}
\end{figure}

\begin{eqnarray}
\begin{array}{lr}
F_2(t)= \frac{1}{2}[ F^{(1)}_1(t)+ F^{(2)}_1(t)],
\\
\\
G_2(t)= \frac{1}{2}[ G^{(1)}_1(t)+ G^{(2)}_1(t)].  \label{fs5f}
\end{array}
\end{eqnarray}
From (\ref{fs5f}) and the information given in the sub-section 3.1
it is evident that when $\alpha_1\neq 0$ and $\alpha_2=0$,
$F^{(2)}_1(t)=0$ and $F^{(1)}_1(t)$ can provide squeezing.
Nevertheless, when $\alpha_j\neq 0$ both of the $F^{(1)}_1(t)$ and
$F^{(2)}_1(t)$ provide nonclassical squeezing. This indicates that
the amount of squeezing produced by the latter case is  greater
than that of the former case.  This fact is remarkable in Fig.
2(a) for given values of the interaction parameters. Moreover, the
comparison between Fig. 1 and Fig. 2(a) shows that for certain
values of the interaction time the nonclassical effects produced
by the two-mode squeezing are greater than those of the
single-mode squeezing. This manifests the role of the correlation
between modes. Now we draw the attention to the general case,
which is plotted in Fig. 2(b). From this figure one can observe
that when $\chi=0$ the nonclassical squeezing occurs in the
$y$-quadrature only (see the long-dashed curve), which is rapidly
increasing as the interaction time evolves. From the short-dashed
curve in Fig. 2(b) one can observe that the Kerr-like medium
causes the nonclassical squeezing occurring periodically in the
$y$-quadrature only with maximum values as those of the
nondegenerate parametric amplifier.
 This indicates in the system under consideration that the kerr-like
 medium decreases the non-classicality produced
by the down-conversion process. Finally, the comparison between
the different curves in the Fig. 2(b) shows that the principal
squeezing produces pure nonclassical effects which for particular
values of the interaction time are greater than those obtained
from the quadrature squeezing.

\subsection{Sum squeezing}
It is worth referring that sum squeezing has been calculated in
nonlinear optics for four-wave sum \cite{sum} and difference
\cite{dif} frequency generation. In this part we investigate sum
squeezing for the system under consideration. In this case  we
have $\hat{B}(t)=\hat{A}_1(t)\hat{A}_2(t)$ and
$\hat{D}(t)=\hat{A}^{\dagger}_1(t)\hat{A}_1(t)
+\hat{A}^{\dagger}_2(t)\hat{A}_2(t)$. We start the discussion with
the case $k=0$, i.e., interaction between two modes via Kerr-like
medium. For this case we obtain
\begin{equation}
F(t)=G(t)=0. \label{su1}
\end{equation}
This means that we have minimum uncertainty sum-squeezing, i.e.
the two-mode-Kerr system cannot generate nonclassical squeezing.
 On the other hand, when $\chi=0$ the system can exhibit squeezing only in the
$y$-quadrature and the associated squeezing factor takes the form

\begin{equation}
G(t)= \frac{ 2\langle\hat{A}_1^{\dagger}(t)\hat{A}_1(t)
\hat{A}_2^{\dagger}(t)\hat{A}_2(t)\rangle -2{\rm
Re}\langle\hat{A}_1^{\dagger 2}(t)\hat{A}^2_2(t)\rangle_{\chi=0}
}{\langle\hat{A}_1^{\dagger}(t)\hat{A}_1(t)\rangle+
\langle\hat{A}_2^{\dagger}(t)\hat{A}_2(t)\rangle}
=\frac{-2(\alpha_1^2+\alpha_2^2+1) S^2 C^2-4\alpha_1 \alpha_2 S
C}{\langle\hat{A}_1^{\dagger}(t)\hat{A}_1(t)\rangle+
\langle\hat{A}_2^{\dagger}(t)\hat{A}_2(t)\rangle}. \label{su2}
\end{equation}
From (\ref{su2}) it is evident that the nonclassical squeezing is
monotonically increasing as the interaction time evolves. This is
obvious from the solid  curve in Fig. 3 for given values of the
interaction parameters.
 Now we draw the attention to the general case in which $\chi\neq 0$
 and $k\neq 0$.
  The  squeezing factors for this case take the forms
\begin{eqnarray}
\begin{array}{lr}
 F(t)= \frac{ 2\langle\hat{A}_1^{\dagger}(t)\hat{A}_1(t)
\hat{A}_2^{\dagger}(t)\hat{A}_2(t)\rangle +2 [{\rm
Re}\langle\hat{A}_1^{\dagger
2}(t)\hat{A}^2_2(t)\rangle_{\chi=0}]\cos(4\chi t) -4[{\rm
Re}\langle\hat{A}_1^{\dagger
}(t)\hat{A}_2(t)\rangle_{\chi=0}]^2\cos^2(2\chi t)
}{\langle\hat{A}_1^{\dagger}(t)\hat{A}_1(t)\rangle+
\langle\hat{A}_2^{\dagger}(t)\hat{A}_2(t)\rangle},
\\
\\
G(t)= \frac{ 2\langle\hat{A}_1^{\dagger}(t)\hat{A}_1(t)
\hat{A}_2^{\dagger}(t)\hat{A}_2(t)\rangle -2 [{\rm
Re}\langle\hat{A}_1^{\dagger
2}(t)\hat{A}^2_2(t)\rangle_{\chi=0}]\cos(4\chi t) -4[{\rm
Im}\langle\hat{A}_1^{\dagger
}(t)\hat{A}_2(t)\rangle_{\chi=0}]^2\sin^2(2\chi t)
}{\langle\hat{A}_1^{\dagger}(t)\hat{A}_1(t)\rangle+
\langle\hat{A}_2^{\dagger}(t)\hat{A}_2(t)\rangle}, \label{su3}
\end{array}
\end{eqnarray}
where {\rm Re} and {\rm Im} stand for the real and imaginary
parts. From (\ref{su2}) and (\ref{su3}) one can realize that the
Kerr-like medium  switches  the nonclassical squeezing
periodically between the two quadratures with maximum values as
those of the nondegenerate parametric amplifier (see the
short-dashed and long-dashed curves in Fig. 3). This  behavior is
different from that of the two-mode squeezing for which
nonclassical effects occur in one quadrature only.
 It is worth mentioning that the principal squeezing for
the sum squeezing is typical as the solid curve in Fig. 3, i.e. it
is  the envelope of the quadrature squeezing.

In conclusion, in this Letter we give for the first time the
competition between the down-conversion and Kerr-like processes
from the point of view of generation nonclassical squeezing. We
prove that generally the amount of the nonclassical squeezing
obtained from each individual process is decreased as a result of
their competition in the Kerr-down conversion system. This has
been shown for different kinds of quadrature squeezing. We have
also suggested more general form for the description of principal
squeezing.

\section*{ Acknowledgment}

J.P. acknowledges the Resarch Project "Measurement and
Information in Optics" MSM 6198959213 and EU Project COST OCP
11.003.

\section*{References}

\end{document}